\documentclass[
superscriptaddress,
twocolumn,
nofootinbib,
amsmath,
amssymb,
aps,
prd
]{revtex4-2}
\usepackage{graphicx}
\usepackage{bm}
\usepackage[colorlinks=true,
       	    linkcolor=blue,
       	    citecolor=blue,
       	    filecolor=blue,
       	    urlcolor=blue,
            ]{hyperref}
\begin{document}

\title{Induced Scattering of Strong Waves in Pair Plasmas}

\author{Masanori Iwamoto}
\email{m-iwamoto@people.kobe-u.ac.jp}
\affiliation{Graduate School of System Informatics, Kobe University, 
1-1 Rokkodai-cho, Nada-ku, Kobe 657-8501  Japan}
\affiliation{Yukawa Institute for Theoretical Physics, Kyoto University, Kitashirakawa-Oiwakecho, 
Sakyo-Ku, Kyoto 606-8502, Japan}

\author{Kunihito Ioka}
\affiliation{Yukawa Institute for Theoretical Physics, Kyoto University, Kitashirakawa-Oiwakecho, 
Sakyo-Ku, Kyoto 606-8502, Japan}

\begin{abstract}

We study induced (stimulated) scattering of linearly polarized, strong electromagnetic waves in pair plasmas, which is crucial for understanding 
the propagation of fast radio bursts (FRBs). Magnetars are the most promising progenitors of FRBs, and FRBs 
propagate through the magnetar wind and successfully escape before being significantly scattered.
We revisit the steady-state solution of linearly polarized electromagnetic waves in pair plasmas 
with arbitrary amplitude, and demonstrate that the nonlinearity is characterized by the nonlinearity parameter $a_0\omega_{pe}/\omega_0$ rather than 
the dimensionless amplitude $a_0$, where $\omega_{pe}$ is the electron plasma frequency and $\omega_0$ is the wave frequency 
in the frame in which the plasma is at rest when the wave fields vanish.
We follow the time evolution of the steady-state solution for the linear regime $a_0\omega_{pe}/\omega_0 \ll 1$ by performing one-dimensional particle-in-cell
simulations, and show that the conventional linear analysis of induced scattering assuming $a_0 \ll 1$ is applicable even for $a_0 > 1$ when the 
Lorentz boost due to the plasma motion in the incident wave is considered. The saturation level is controlled by
$a_0\omega_0/\omega_{pe}$, which corresponds to the ratio of the wave energy to the plasma energy, and the incident wave is hardly scattered 
for $a_0\omega_0/\omega_{pe} \gg 1$. We discuss the application of our results to FRBs.

\end{abstract}

\maketitle
	
\section{\label{sec:intro} Introduction}

Strong electromagnetic waves are ubiquitous in the universe,
with fast radio bursts (FRBs) being the most prominent example.
FRBs are millisecond-long bright flashes of radio waves, mostly from extragalactic
distances \cite{Lorimer2007, Petroff2022,Zhang2023}.
Some FRBs are known to burst repeatedly, 
and such repeating FRBs often show the high degree of linear polarization \cite{Masui2015,Michilli2018,Luo2020}. 
Magnetars are the most likely progenitors of repeating FRBs \cite{Andersen2020,Bochenek2020,Mereghetti2020,Ridnaia2021,Li2021}.
The mechanism by which FRBs are generated in magnetars remains a topic of debate,
and numerous theoretical models have been proposed, including
coherent curvature radiation in the magnetosphere \cite{Katz2014,Kumar2017,Katz2018,Lu2018,Lu2020,Wang2022}, an expanding fireball 
\cite{Ioka2020, Wada2023}, fast magnetosonic waves via reconnection \cite{Lyubarsky2020,Lyubarsky2021,Mahlmann2022}, and
relativistic magnetized shocks in pair (electron-positron) plasmas 
\cite{Lyubarsky2014,Waxman2017,Beloborodov2017, Beloborodov2020,Plotnikov2019,Sironi2021,Vanthieghem2025} 
or electron-ion plasmas \cite{Metzger2019, Margalit2019, Margalit2020,Iwamoto2024}, among others.
In all of these models, FRBs must propagate through plasmas surrounding their sources and successfully escape. 
The radio waves are 
strong in the sense that the normalized wave electric field, called the strength parameter, exceeds unity,
\begin{equation}
	a_0 = \frac{eE_0}{m_ec\omega_0} > 1
\end{equation}
for distances from the source $R \lesssim 10^{13}$cm \cite{Luan2014},
where $e$ and $m_e$ are the electron charge and mass, $E_0$ is the amplitude of the wave electric field, 
and $\omega_0$ is the wave frequency.
Such strong waves inevitably suffer from induced (or stimulated) scattering 
\footnote{Induced scattering and stimulated scattering are essentially the same process and are terms often used interchangeably.} 
\cite{Lyubarsky2008,Lyubarsky2019}, which could hinder their propagation and constrain the emission region
\cite{Beloborodov2021,Beloborodov2022,Beloborodov2024,Sobacchi2024a,Sobacchi2025,Nishiura2025a,Nishiura2025b,Kamijima2026,Nishiura2026}.
On the other hand, the effect of induced scattering on FRB propagation remains controversial, 
especially for $a_0 > 1$ \cite{Qu2022,Lyutikov2024,Qu2024}.

One of the main challenges in studying induced scattering in FRBs lies in the analytical intractability of  
the self-consistent equations for linearly polarized electromagnetic plane waves with arbitrary amplitude, 
even in unmagnetized plasmas \cite{Akhiezer1956,Kaw1970, Max1971,Max1973b,Clemmow1974,Kennel1976}.
Although a similar problem has been addressed in the context of pulsars \cite{Kennel1973,Arka2012,Mochol2013},
it has received little attention in FRBs. Previous analyses of linearly polarized electromagnetic waves in pair plasmas 
\cite{Ghosh2022,Iwamoto2023} were limited to the regime $a_0 \ll 1$, where relativistic effects are negligible and 
the steady-state solution can be expressed solely in terms of elementary functions. Recently, Ref. 
\cite{Sobacchi2024b} studied the stability of wave packets and 
demonstrated that nonlinear effects remain negligible
when the nonlinearity parameter is sufficiently smaller than unity,
\begin{equation}
	a_0 \frac{\omega_{pe}}{\omega_0} \ll 1,
\end{equation}
where $\omega_{pe}$ 
is the plasma frequency and $\omega_0$ is the wave frequency 
in the frame in which the plasma is at rest when the wave fields vanish.
This behavior arises because, in this regime, the plasma current can be approximated as a linear function of the vector potential.
Since the FRB frequency is sufficiently high $\omega_0 \gg \omega_{pe}$ \cite{Sobacchi2023}, the condition 
$a_0 \omega_{pe}/\omega_0 \ll 1$ can be satisfied and the linear treatment can remain valid for $a_0 > 1$.
This result implies that the previous studies on induced scattering can be extrapolated to 
the regime $a_0 > 1$ when $a_0 \omega_{pe}/\omega_0 \ll 1$.

In this paper, we revisit the self-consistent equations of linearly polarized electromagnetic waves with arbitrary 
amplitude and study the steady-state solution. 
We also perform particle-in-cell (PIC) simulation and 
follow the time evolution of the steady-state solution for $a_0 \omega_{pe}/\omega_0 \ll 1$.
This paper is organized as follows.
In Section \ref{sec:analytical}, we derive the self-consistent equations following 
the previous studies and investigate the parameter dependence of the solution.
Section \ref{sec:simulation} describes our simulation results.
We compare them with the linear analysis of induced scattering and discuss the saturation.
In Section \ref{sec:discuss}, we apply our results to FRBs.
We finally summarize our results in Section \ref{sec:summary}.

\section{\label{sec:analytical} Analytical Formulation}
\subsection{Basic Equations}
We derive the self-consistent equations for linearly polarized electromagnetic waves with arbitrary amplitude 
following previous works \cite{Akhiezer1956,Kaw1970,Max1971,Max1973b,Clemmow1974,Kennel1976,Arka2012,Mochol2013}.
Throughout this paper, unless explicitly specified, we use the frame
in which the plasma would be at rest in the absense of the incident electromagnatic wave and refer to this frame as the laboratory frame.
Basic equations are the relativistic, cold two-fluid equations
and Maxwell equations in the laboratory frame,
\begin{eqnarray}
  \frac{\partial}{\partial t} (\gamma_\pm n_\pm)+ \bm{\nabla} \cdot (\gamma_\pm n_\pm \bm{v_\pm})= 0, \\
  \frac{\partial \bm{u_\pm}}{\partial t} + (\bm{v_\pm} \cdot \bm{\nabla})\bm{u_\pm} 
  = \pm \frac{e}{m_e c} \left[\bm{E} + \frac{\bm{v_\pm}}{c} \times \bm{B}\right], \\
  \bm{\nabla} \cdot \bm{E} = 4\pi \rho, \\
  \bm{\nabla} \cdot \bm{B} = 0, \\
  \bm{\nabla} \times \bm{E} = -\frac{1}{c}\frac{\partial \bm{B}}{\partial t}, \\
  \bm{\nabla} \times \bm{B} = \frac{4\pi}{c} \bm{j} + \frac{1}{c}\frac{\partial \bm{E}}{\partial t}, \\
  \rho = e(\gamma_+n_+ - \gamma_-n_-), \\
  \bm{j} = e(\gamma_+n_+\bm{v_+} - \gamma_-n_-\bm{v_-}),
\end{eqnarray}
where the plus (minus) index denotes positron (electron), $\bm{v_{\pm}}$ is the particle three velocity, 
$\bm{u_{\pm}}= \gamma_\pm \bm{v_{\pm}}/c$ is the four velocity normalized by the speed of light $c$, and
$\gamma_\pm=1/\sqrt{1-|\bm{v_{\pm}}|^2/c^2}=\sqrt{1+|\bm{u_{\pm}}|^2}$ is the particle Lorentz factor.
Unless otherwise stated, all unprimed fluid and field quantities are evaluated in the laboratory frame,
except that $n_{\pm}$ denotes the proper density measured in the local plasma rest frame.
The density measured in the laboratory frame is therefore $\gamma_{\pm}n_{\pm}$.
We consider a monochromatic plane electromagnetic wave propagating in the $x$ direction with a superluminal phase velocity,
linearly polarized in the $y$ direction. We assume that all physical quantities can be expressed as a function of the phase,
\begin{equation}
	\phi= \omega_0 t - k_0 x,
\end{equation}
where $k_0$ and $\omega_0$ denote the wavenumber and frequency of the incident
electromagnetic wave in the laboratory frame, respectively.
We consider a superluminal wave satisfying $\omega_0/k_0 > c$.
The basic equations are then written as 
\begin{eqnarray}
	\label{eq:continuity}
	\frac{{\rm d}}{{\rm d}\phi} [(\gamma_\pm -\beta_gu_{x\pm})n_{\pm}] = 0, \\
	(\gamma_\pm-\beta_gu_{x\pm}) \frac{{\rm d}u_{x\pm}}{{\rm d}\phi} = \pm \frac{eB_zu_{y\pm}}{m_ec\omega_0}, \\
	(\gamma_\pm-\beta_gu_{x\pm}) \frac{{\rm d}u_{y\pm}}{{\rm d}\phi} 
	= \pm \left(\frac{eE_y\gamma_\pm}{m_ec\omega_0}-\frac{eB_zu_{x\pm}}{m_ec\omega_0}\right), \\
	\frac{{\rm d} B_z}{{\rm d}\phi} = \beta_g \frac{{\rm d} E_y}{{\rm d}\phi}, \\
	\label{eq:ampere}
	\frac{{\rm d} E_y}{{\rm d}\phi} = -\frac{4\pi ec\gamma_g^2(n_+u_{y+} - n_-u_{y-})}{\omega_0}, 
\end{eqnarray}
where
\begin{eqnarray}
	\beta_g = \frac{c k_0}{\omega_0}, \\
	\gamma_g = \frac{1}{\sqrt{1-\beta_g^2}}= \frac{\omega_0}{\sqrt{\omega_0^2-c^2k_0^2}}.
\end{eqnarray}
The parameter $c\beta_g$ corresponds to the velocity of a reference frame moving 
relative to the laboratory frame, 
in which the spatial dependence of both particle and field variables vanishes \cite{Clemmow1974}.
This velocity can be interpreted as the group velocity of the wave \cite{Mochol2013}.
It is convenient to introduce the normalized electric field $y=E_y/E_0$,
where the wave electric field $E_y$ is assumed to take the maximum value $E_0$ at $\phi = 0$, i.e. $y=1$ at $\phi=0$.
Since we can set $\gamma \equiv \gamma_+ = \gamma_-$, $u_x \equiv u_{x+}=u_{x-}$, $u_y \equiv u_{y+}=-u_{y-}$, 
and $n \equiv n_+=n_-$ \cite{Kennel1976}, $\gamma$, $u_x$, $u_y$, and $n$ are expressed in terms of $y$,  
\begin{eqnarray}
	\label{eq:gam}
	\gamma = 1 + \frac{\alpha a_0^2}{2} (1-y^2), \\
	\label{eq:ux}
	u_x = \frac{\alpha \beta_g a_0^2}{2} (1-y^2), \\
	\label{eq:uy}
	u_y = a_0 \int_0^\phi y {\rm d}\phi, \\
	\label{eq:n}
	n =  n_0 \left[ 1+\frac{2}{q}(1-y^2) \right]^{-1},
\end{eqnarray}
where 
\begin{eqnarray}
	\label{eq:alpha}
	\alpha = \frac{\omega_0^2}{2\gamma_g^2\omega_{pe}^2}= \frac{\omega_0^2-c^2k_0^2}{2\omega_{pe}^2}, \\
	\label{eq:q}
	q = \frac{4\gamma_g^2}{\alpha a_0^2}.
\end{eqnarray}
The parameter $a_0$ is the strength parameter defined above and is Lorentz invariant.
The value of $\gamma$ at $\phi = 0$ can be chosen arbitrarily
\cite{Clemmow1974}. Here, we choose $\gamma=1$ at $\phi = 0$, so that the
proper density at this phase is equal to the reference density $n_0$ of the
plasma in the absence of the electromagnetic wave.
The reference electron plasma frequency is defined from this density as
\begin{equation}
	\omega_{pe} = \sqrt{\frac{4\pi n_0 e^2}{m_e}}.
\end{equation}
Throughout this paper, $\omega_{pe}$ denotes this reference plasma frequency
and is not redefined in Lorentz-transformed frames. This definition of $n_0$
and $\omega_{pe}$ is consistent with that used in Ref. \cite{Sobacchi2024b}.

Eq. \ref{eq:uy} represents the conservation law of canonical momentum.
The parameters $\alpha$ and 
$\beta_g$ originate from the plasma dispersion, and
Eqs. \ref{eq:gam}, \ref{eq:ux}, and \ref{eq:uy} describe the motion of test particles in
transverse electromagnetic waves when these factors are neglected \cite{Gunn1971}.
The normalized electric field $y$ is determined from the differential equation
with the boundary condition $y=1$ at $\phi=0$
\footnote{
Eq. \ref{eq:y} is equivalent to Eq. 83 of Ref.
\cite{Clemmow1974}, but is written with a different dependent variable and
frame convention. Ref. \cite{Clemmow1974} writes the equation in the frame
where the spatial dependence vanishes, and uses the transverse four-velocity
component $\eta$ ($u_y$ in our notation). Here, we use the normalized
electric field $y=E_y/E_0$, which is related to $u_y$ by Eq. \ref{eq:uy}.
},
\begin{equation}
	\label{eq:y}
	\frac{\alpha^2a_0^2}{\gamma_g^2}\left( \frac{{\rm d}y}{{\rm d}\phi}\right)^2
	=\frac{(1-y^2)(1-y^2+q)}{(1-y^2+q/2)^2}.
\end{equation}
The dispersion relation follows from the fact
that the phase shifts by $\pi/2$ after a quarter-cycle,
\begin{equation}
	\int^1_0 \frac{1}{\left| {\rm d}y /{\rm d}\phi\right|} {\rm d}y = \frac{\pi}{2}.
\end{equation}
By substituting Eq. \ref{eq:y} into this dispersion relation, one can find
\begin{equation}
	\label{eq:disp}
	\frac{\alpha a_0}{\gamma_g } \frac{2E(m)-(1-m)K(m)}{2\sqrt{m}}=\frac{\pi}{2},
\end{equation}
where $m = 1/(1+q)$ and $0<m<1$. $K(m)$ and $E(m)$ are the complete elliptic 
integral of the first and second kind with modules $m$, 
\begin{eqnarray}
	K(m) &=& \int_{0}^{1}\frac{1}{\sqrt{(1-y^2)(1-my^2)}}{\rm d}y \nonumber \\
	&=& \int_{0}^{\frac{\pi}{2}}\frac{1}{\sqrt{1-m\sin^2\theta}}{\rm d}\theta, \\
	E(m) &=& \int_{0}^{1}\sqrt{\frac{1-my^2}{1-y^2}}{\rm d}y \nonumber \\
	&=& \int_{0}^{\frac{\pi}{2}}\sqrt{1-m\sin^2\theta} {\rm d}\theta.
\end{eqnarray}
Our final set of equations consists of Eqs. \ref{eq:gam}, \ref{eq:ux}, \ref{eq:uy}, 
\ref{eq:n}, \ref{eq:y}, and \ref{eq:disp} for a given normalized amplitude $a_0$ and 
frequency $\omega_0/\omega_{pe}$.

\subsection{Steady-State Solution}

We demonstrate that the steady-state solution of linearly polarized
electromagnetic waves is controlled by the nonlinearity parameter
$a_0\omega_{pe}/\omega_0$ and that the linear approximation holds for
$a_0\omega_{pe}/\omega_0 \ll 1$. 
Here, $\omega_0$ is measured in the
laboratory frame, while $\omega_{pe}$ is the reference plasma frequency
defined from $n_0$. 
Therefore, $a_0\omega_{pe}/\omega_0$ is not Lorentz invariant and it should be 
understood as the nonlinearity parameter defined in the laboratory frame.

Considering Eqs. \ref{eq:alpha} and \ref{eq:q}, 
the coefficient $\alpha a_0/\gamma_g$ is expressed as
\begin{eqnarray}
	\label{eq:fac}
	\frac{\alpha a_0}{\gamma_g} = \left(\frac{32}{q^3}\right)^{\frac{1}{4}}\left(a_0\frac{\omega_{pe}}{\omega_0}\right)^{-\frac{1}{2}}.
\end{eqnarray}
By substituting this into Eq. \ref{eq:disp} and using $q=(1-m)/m$, we obtain
\begin{equation}
	\label{eq:disp2}
	\frac{m\left[2E(m)-(1-m)K(m)\right]^4}{(1-m)^3}=\frac{1}{2}\left(\frac{\pi}{2}\right)^4\left(a_0\frac{\omega_{pe}}{\omega_0}\right)^2,
\end{equation}
indicating that $m$ (and thus $q$) can be expressed as a function of the nonlinearity parameter $a_0\omega_{pe}/\omega_0$.
For $m \ll 1$, 
$K(m)$ and $E(m)$ can be expanded as
\begin{eqnarray}
	K(m) \simeq \int_{0}^{\frac{\pi}{2}}\left(1+\frac{1}{2}m\sin^2\theta\right){\rm d}\theta = \frac{\pi}{2}\left(1+\frac{m}{4}\right), \\
	E(m) \simeq \int_{0}^{\frac{\pi}{2}}\left(1-\frac{1}{2}m\sin^2\theta\right){\rm d}\theta = \frac{\pi}{2}\left(1-\frac{m}{4}\right).
\end{eqnarray}
By substituting these into Eq. \ref{eq:disp2} and keeping the lowest-order of $m$,
we obtain
\begin{equation}
	m \simeq \frac{1}{2}\left(a_0\frac{\omega_{pe}}{\omega_0}\right)^2.
\end{equation}
The validity condition $m \ll 1$ now becomes
\begin{equation}
	a_0\frac{\omega_{pe}}{\omega_0} \ll 1.
\end{equation}
On the other hand, for $1-m \ll 1$, $K(m)$ and $E(m)$ can be expanded as \cite{Cody1965},
\begin{eqnarray}
	K(m) &\simeq& \ln{4}-\frac{1}{2}\ln{(1-m)}, \\
	E(m) &\simeq& 1+\frac{\ln{4}-1}{2}(1-m)-\frac{1}{4}(1-m)\ln{(1-m)}. \ \ \ \ \
\end{eqnarray}
Keeping the lowest-order of $1-m$, we obtain
\begin{equation}
	1-m \simeq \frac{8}{\pi^{\frac{4}{3}}}\left(a_0\frac{\omega_{pe}}{\omega_0}\right)^{-\frac{2}{3}}.
\end{equation}
The validity condition $1-m \ll 1$ can be rewritten as
\begin{equation}
	a_0\frac{\omega_{pe}}{\omega_0} \gg 1.
\end{equation}

\begin{figure}[htbp]
	\includegraphics[width=8cm]{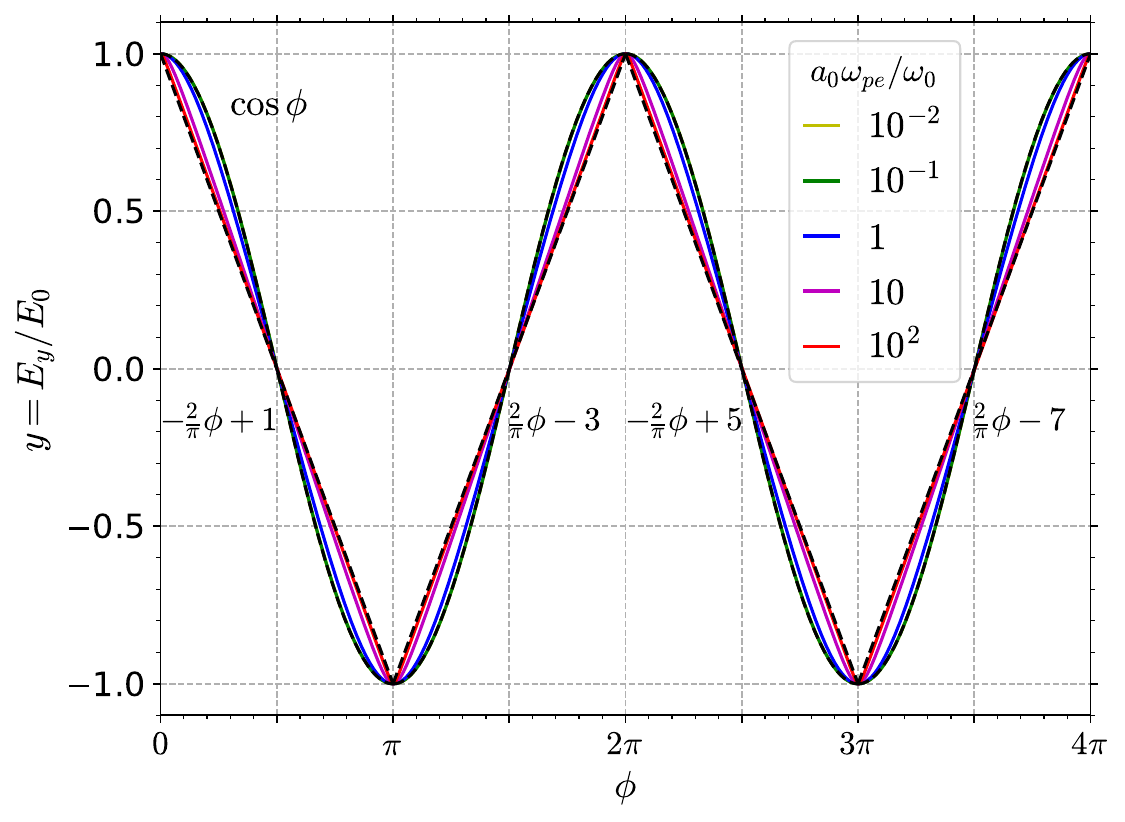}
	\caption{The wave electric field $y=E_y/E_0$ as a function of the phase $\phi$.
	The color indicates the numerical solutions for different values of the nonlinearity parameter $a_0\omega_{pe}/{\omega_0}$.
	The black dashed lines correspond to the analytical asymptotic solutions
	for $a_0\omega_{pe}/{\omega_0} \ll 1$ and $a_0\omega_{pe}/{\omega_0} \gg 1$.}
	\label{fig:Ey}
\end{figure}

One can find the wave electric field $y$ after determining the solution of Eq. \ref{eq:disp2}. 
By substituting Eq. \ref{eq:fac} into Eq. \ref{eq:y}, we obtain
\begin{equation}
	\label{eq:y2}
	\left( \frac{{\rm d}y}{{\rm d}\phi}\right)^2 = a_0\frac{\omega_{pe}}{\omega_0} \frac{q^{\frac{3}{2}}}{4\sqrt{2}} \frac{(1-y^2)(1-y^2+q)}{(1-y^2+q/2)^2}.
\end{equation}
This demonstrates that $y$ is well-characterized by the nonlinearity parameter $a_0\omega_{pe}/{\omega_0}$.
For $a_0\omega_{pe}/{\omega_0} \ll 1$ (i.e. $q=(1-m)/m \gg 1$),
we consider the zeroth order of $a_0\omega_{pe}/{\omega_0}$ and 
Eq. \ref{eq:y2} can be written as
\begin{equation}
	\frac{{\rm d}y}{{\rm d}\phi}= \pm \sqrt{1-y^2}.
\end{equation}
Note that $y^2 \leq 1$ is satisfied by definition.
This derivative equation is easily solved for the boundary condition $y=1$ at $\phi=0$ ,
\begin{equation}
	\label{eq:y_lin}
	y = \cos{\phi}.
\end{equation}
For $a_0\omega_{pe}/{\omega_0} \gg 1$ (i.e. $q=(1-m)/m \ll 1$),
the zeroth-order equation is expressed as
\begin{equation}
	\frac{{\rm d}y}{{\rm d}\phi}=\pm\frac{2}{\pi}.
\end{equation}
The periodic solution is given by
\begin{equation}
	\label{eq:y_nl}
	y = 
	\begin{cases}
		-\frac{2}{\pi}\phi+4n+1 & \text{if $2n\pi \leq \phi \leq (2n+1)\pi$,} \\
		\frac{2}{\pi}\phi-4n-3  & \text{if $(2n+1)\pi \leq \phi \leq 2(n+1)\pi$,} 
	\end{cases}
\end{equation}
where $n=0,1,2,...$ is an integer.
We have numerically determined $m$ from Eq. \ref{eq:disp2} and solved Eq. \ref{eq:y2}.
Fig. \ref{fig:Ey} shows the numerical solutions for $y$ at various values of $a_0\omega_{pe}/{\omega_0}$:
$10^{-1}$ (yellow), $1$ (green), $10$ (blue), $10^2$ (magenta), and $10^3$ (red).
The steady-state solution asymptotically approaches the linear one $y = \cos \phi$ 
for $a_0\omega_{pe}/{\omega_0} \ll 1$. For $a_0\omega_{pe}/{\omega_0} \gg 1$, the wave electric 
field has a sawtooth-like profile, which is consistent with previous studies \cite{Max1971,Max1973b}

\begin{figure}[htbp]
	\includegraphics[width=8cm]{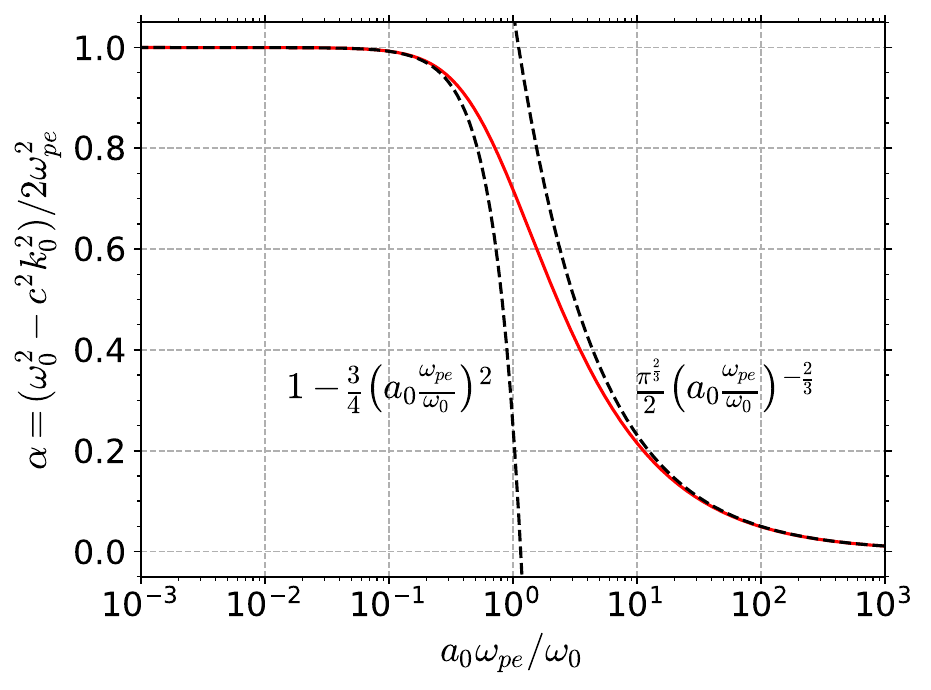}
	\caption{The parameter $\alpha=(\omega_0^2-c^2k_0^2)/2\omega_{pe}^2$ 
	as a function of the nonlinearity parameter $a_0\omega_{pe}/{\omega_0}$ (red solid line). 
	The black dashed lines correspond to the asymptotic solutions for $a_0\omega_{pe}/{\omega_0} \ll 1$ 
	and $a_0\omega_{pe}/{\omega_0} \gg 1$.}
	\label{fig:disp}
\end{figure}

The parameter $\alpha$ can be determined from the solution of Eq. \ref{eq:disp2} as well.
Considering $a_0^2/\gamma_g^2 =  4/q\alpha$ and $q=(1-m)/m$,
Eq. \ref{eq:disp} is rewritten as
\begin{equation}
	\alpha =  \left(\frac{\pi}{2}\right)^2\frac{1-m}{[2E(m)-(1-m)K(m)]^2}.
\end{equation}
This demonstrates that $\alpha$ also depends solely on the nonlinearity parameter $a_0\omega_{pe}/{\omega_0}$. 
We retain the lowest-order of $m$ when $a_0\omega_{pe}/{\omega_0} \ll 1$ (i.e. $m \ll 1$),
\begin{eqnarray}
	\alpha \simeq 1-\frac{3}{2}m \simeq 1-\frac{3}{4}\left(a_0\frac{\omega_{pe}}{\omega_0}\right)^2,
\end{eqnarray}
and $1-m$ when $a_0\omega_{pe}/{\omega_0} \gg 1$ (i.e. $1-m \ll 1$), 
\begin{eqnarray}
	\alpha \simeq \frac{1}{4} \left(\frac{\pi}{2}\right)^2(1-m) 
	\simeq \frac{\pi^{\frac{2}{3}}}{2}\left(a_0\frac{\omega_{pe}}{\omega_0}\right)^{-\frac{2}{3}}.
\end{eqnarray}
Fig. \ref{fig:disp} shows the numerical solution $\alpha$ as a function of $a_0\omega_{pe}/{\omega_0}$ 
in the red solid line and the asymptotic solutions in the dashed black lines.
In the limit $a_0\omega_{pe}/{\omega_0} \ll 1$, the parameter $\alpha$ asymptotically approaches unity $\alpha=1$,
thereby recovering the linear dispersion relation $\omega_0^2 = 2\omega_{pe}^2+c^2k_0^2$.

The Lorentz factor $\gamma_g$ can be determined from Eq. \ref{eq:alpha} after the dispersion relation $\alpha$ is obtained.
One can find
\begin{equation}
	\gamma_g \simeq \frac{\omega_0}{\sqrt{2}\omega_{pe}}\left[1+\frac{3}{8}\left(a_0\frac{\omega_{pe}}{\omega_0}\right)^2\right],
\end{equation}
for  $a_0\omega_{pe}/{\omega_0} \ll 1$, and
\begin{equation}
	\gamma_g \simeq \frac{\omega_0}{\pi^{\frac{1}{3}}\omega_{pe}}\left(a_0\frac{\omega_{pe}}{\omega_0}\right)^{\frac{1}{3}},
\end{equation}
for  $a_0\omega_{pe}/{\omega_0} \gg 1$.
When $a_0\omega_{pe}/{\omega_0} \ll 1$, the Lorentz factor $\gamma_g$ is approximately equal to $\omega_0/\sqrt{2}\omega_{pe}$ at 
the lowest order. This is the same as the group velocity Lorentz factor of the wave packet in Ref. \cite{Sobacchi2024b},
indicating that our treatment is consistent with the wave packet analysis in the linear regime.

In the vacuum limit, $\omega_{pe}/\omega_0\to0$, which corresponds to
$a_0\omega_{pe}/\omega_0\to0$ for fixed $a_0$, the plasma feedback on the
electromagnetic wave is absent, and the waveform is prescribed as
$y=\cos\phi$. In this limit, $\alpha=1$ and $\beta_g=1$, and Eqs.
\ref{eq:gam}--\ref{eq:uy} reduce to
\begin{align}
	\gamma &= 1 + \frac{a_0^2}{2}\sin^2\phi, \\
	u_x &= \frac{a_0^2}{2}\sin^2\phi, \\
	u_y &= a_0\sin\phi .
\end{align}
For $a_0\ll1$, these equations further give the usual nonrelativistic quiver
orbit.

The nonlinear feedback of the plasma on the electromagnetic wave is mediated by the plasma current,
\begin{equation}
	j_y = 2 e n c u_y.
\end{equation}
For $a_0\omega_{pe}/\omega_0 \ll 1$ (i.e., $q \gg 1$), Eq. \ref{eq:n} gives $n \simeq n_0$ at leading order. The conservation of canonical momentum (Eq. \ref{eq:uy} )
then implies $u_y = eA_y/m_ec$, where 
$A_y$ is the vector potential, so that the current reduces to
\begin{equation}
	j_y \simeq 2 e n_0 c u_y \propto A_y.
\end{equation}
Hence, the source term in Maxwell's equation remains linear in the wave amplitude, and 
the plasma response reduces to the test-particle limit. In particular, no additional 
amplitude-dependent coupling is generated between the wave and the plasma, so the 
waveform  does not undergo nonlinear distortion. Nonlinear feedback is therefore negligible for 
$a_0\omega_{pe}/\omega_0 \ll 1$. We thus conclude that the steady-state solution of 
linearly polarized electromagnetic waves is controlled by the nonlinearity parameter 
$a_0\omega_{pe}/\omega_0$, and that the linear treatment remains valid in this regime, 
consistent with Ref. \cite{Sobacchi2024b}.

\subsection{Induced Scattering} \label{subsec:sbs}
Induced scattering can be understood as a parametric instability, which 
has been studied through the stability analysis of plasma waves \cite{Drake1974,Forslund1975,Kruer1988,Matsukiyo2003,Ishizaki2024}. 
Linearly polarized electromagnetic waves traveling through unmagnetized pair plasmas 
are subject to stimulated Brillouin scattering (SBS)
\footnote{We refer to this process as ``stimulated'' Brillouin scattering 
rather than ``induced'' Brillouin scattering, as the former term is more widely used in the literature.}. 
SBS is often referred to as 
induced Compton scattering when kinetic effects are important, 
as is always the case for unmagnetized pair plasmas \cite{Schluck2017}.
Previous studies \cite{Ghosh2022,Iwamoto2023} evaluated the linear growth rate of SBS $\Gamma={\rm Im} \ \omega_1$ and
the wavenumber of the scattered wave $k_1$ under the assumption that the incident wave is weak $a_0 \ll 1$ 
and the plasma temperature is non-relativistic.
We assume that the SBS operates in a frame where the averaged longitudinal moment vanishes, 
and that the linear analysis of SBS remains valid in this center-of-momentum frame.
The drift velocity $v_D$ of the center-of-momentum  frame relative to the laboratory frame is given by \cite{Sarachik1970}
\begin{equation}
	\label{eq:vD}
	v_D= \frac{c\langle u_x \rangle}{\langle \gamma \rangle} = \frac{c\beta_g \alpha a_0^2(1-\langle y^2 \rangle)/2}{1+\alpha a_0^2(1-\langle y^2 \rangle)/2},
\end{equation}
with Eqs. \ref{eq:gam} and \ref{eq:ux}.
This dirft velocity depends not only on the nonlinearity parameter $a_0\omega_{pe}/\omega_0$ but also on $a_0$ itself.
Here $\langle \cdots \rangle$ denotes an average taken over a time interval much longer than the wave period but much shorter than the SBS growth timescale.
{Note that this drift velocity is different from the averaged three velocity 
$c\langle u_x / \gamma \rangle$.}
One can find $\langle y^2 \rangle \simeq 1/2$ for $a_0\omega_{pe}/\omega_0 \ll 1$ and  $\langle y^2 \rangle \simeq 1/3$ for $a_0\omega_{pe}/\omega_0 \gg 1$
with Eqs. \ref{eq:y_lin} and \ref{eq:y_nl}, respectively.
In the center-of-momentum frame, we assume that the maximum growth rate $\Gamma_{\mathrm{max}}^{\prime}$ and the corresponding wavenumber $k_{1,\mathrm{max}}^{\prime}$ 
can be derived from the linear theory \cite{Ghosh2022,Iwamoto2023},
\begin{eqnarray}
	\frac{\Gamma_{\mathrm{max}}^{\prime}}{\omega_0^{\prime}} &=& \sqrt{\frac{\pi}{32e}}\left(a_0\frac{\omega_{pe}}{\omega_0^{\prime}}\right)^2\frac{1}{\beta_{th0}^2},\\
  	\frac{k_{1,\mathrm{max}}^{\prime}}{k_0^{\prime}} &=& -(1-2\beta_{th0}),
\end{eqnarray}
for the weak coupling regime $\beta_{th0} \gg (a_0\omega_{pe}/\omega_0^{\prime})^{2/3}$. 
Here the primed quantities are defined in the center-of-momentum frame and $\beta_{th0}=\sqrt{k_BT_0/m_ec^2} \ll 1$ is the initial thermal velocity
defined by the proper temperature $T_0$. Note that $a_0$ is Lorentz invariant and 
{$\omega_{pe}$ is defined by the reference density $n_0$.}
The negative wavenumber indicates the backward scattering.
For $\beta_{th0} \ll (a_0\omega_{pe}/\omega_0^{\prime})^{2/3}$, which is the strong coupling regime \cite{Forslund1975},
they are given by
\begin{eqnarray}
	\frac{\Gamma_{\mathrm{max}}^{\prime}}{\omega_0^{\prime}} &=& \frac{\sqrt{3}}{2}\left(a_0\frac{\omega_{pe}}{\omega_0^{\prime}}\right)^{\frac{2}{3}},\\
  	\frac{k_{1,\mathrm{max}}^{\prime}}{k_0^{\prime}} &=& -1,
\end{eqnarray} 
The strong-coupling scaling
$(a_0\omega_{pe}/\omega_0')^{2/3}$ is closely connected to the scaling recently
discussed for pair plasmas by Ref. \cite{Lyutikov2025}.
These results indicate that the SBS is well characterized by the nonlinearity parameter in the center-of-momentum frame.
We also assume that the linear theory can be extrapolated to the regime $a_0 > 1$ as long as the nonlinearity parameter is 
sufficiently small, $a_0\omega_{pe}/\omega_0 \ll 1$.
Since the incident wave strongly drives the plasma in the $+x$ direction for $a_0 > 1$ 
and the Lorentz boost effect due to $v_D$ is not negligible (see Eq. \ref{eq:vD}), the 
maximum growth rate $\Gamma_{\mathrm{max}}$  in the laboratory frame should satisfy 
the Lorentz transformation \cite{Hasegawa1978}
\begin{equation}
	\label{eq:gamlab}
	\Gamma_{\mathrm{max}}^{\prime} =\frac{\Gamma_{\mathrm{max}}}{\gamma_D(1-\beta_g\beta_D)},
\end{equation}
where 
\begin{eqnarray}
	\beta_D= \frac{v_D}{c}, \\
	\gamma_D= \frac{1}{\sqrt{1-\beta_D^2}}.
\end{eqnarray}
Here we have assumed that the back-scattered wave has a phase velocity of $\omega_1/k_1 \simeq -\omega_0/k_0$ in the laboratory frame.
The wavenumber of the back-scattered wave satisfies
\begin{equation}
	\label{eq:klab}
  	k_{1,\mathrm{max}}^{\prime} = \gamma_D\left(1+\frac{\beta_D}{\beta_g}\right) k_{1,\mathrm{max}}.
\end{equation}
The Lorentz transformation of the frequency and wavenumber of the incident wave is given by
\begin{eqnarray}
	\omega_0^{\prime} = \gamma_D(1-\beta_g\beta_D)\omega_0,\\
	k_0^{\prime} = \gamma_D\left(1-\frac{\beta_D}{\beta_g}\right) k_0.
\end{eqnarray}	
We finally obtain the maximum growth rate and the wavenumber of the scattered wave in the laboratory frame as
\begin{eqnarray}
	\label{eq:gamlab1}
	\frac{\Gamma_{\mathrm{max}}}{\omega_0} &=& \sqrt{\frac{\pi}{32e}}\left(a_0\frac{\omega_{pe}}{\omega_0}\right)^2\frac{1}{\beta_{th0}^2},\\
	\label{eq:klab1}
  	\frac{k_{1,\mathrm{max}}}{k_0} &=& -\frac{\beta_g-\beta_D}{\beta_g+\beta_D}(1-2\beta_{th0}),
\end{eqnarray}
for the weak coupling regime and 
\begin{eqnarray}
	\label{eq:gamlab2}
	\frac{\Gamma_{\mathrm{max}}}{\omega_0} &=& \frac{\sqrt{3}}{2}\left(a_0\frac{\omega_{pe}}{\omega_0}\right)^{\frac{2}{3}}[\gamma_D(1-\beta_g\beta_D)]^{\frac{4}{3}},\\
	\label{eq:klab2}
  	\frac{k_{1,\mathrm{max}}}{k_0} &=& -\frac{\beta_g-\beta_D}{\beta_g+\beta_D},
\end{eqnarray}
for the strong coupling regime.
When $a_0\omega_{pe}/{\omega_0} \ll 1$ and $\omega_0 \simeq ck_0$ are satisfied, one can find 
\begin{eqnarray}
	\label{eq:gam_D2}
	\frac{\beta_g-\beta_D}{\beta_g+\beta_D} &\simeq& \left(1+\frac{a_0^2}{2}\right)^{-1}, \\
	\label{eq:gam_D}
	\gamma_D(1-\beta_g\beta_D) &\simeq& \left(1+\frac{a_0^2}{2}\right)^{-\frac{1}{2}}.
\end{eqnarray}
These equations indicate that both the growth rate and the wavenumber of the scattered wave in the simulation frame depend not only on the nonlinearity parameter 
$a_0\omega_{pe}/\omega_0$ but also on the incident wave amplitude $a_0$ through $\beta_D$ and $\gamma_D$, except for the maximum growth rate in the weak coupling case 
(Eq. \ref{eq:gamlab1}). In the above analysis, we neglect relativistic mass effects associated with the rapid quiver motion of the particles. 
This approximation is justified in the regime $a_0\omega_{pe}/\omega_0 \ll 1$, where the wave dynamics is well approximated by the linear solution 
and the SBS growth timescale is much longer than the incident wave period, i.e., $\Gamma_{\mathrm{max}} \ll \omega_0$, so that a secular description is well defined. 
When this condition is not satisfied, relativistic mass effects become important, which can significantly reduce the instability 
\cite{Lyubarsky2019}. We perform numerical simulations to test the above analysis.

\section{\label{sec:simulation} Kinetic Simulations}

\subsection{\label{subsec:simulation} Simulation Setup}
We use a fully kinetic PIC code, WumingPIC \cite{Matsumoto2024}.
The simulation setup is based on previous studies \cite{Ghosh2022,Iwamoto2023} 
and the physics of SBS is fully captured. 
We consider one-dimensional (1D) spatial domain in the $x$ direction and the periodic boundary condition is 
applied for both particles and fields. 
All three components of fields and velocities are tracked in our simulations.
The simulation frame corresponds to the laboratory frame in Section \ref{sec:analytical}.
We initially introduce a strong wave as described below.
The wave amplitude and frequency are given as
$(a_0,\omega_0/\omega_{pe}) = (0.05,5)$, $(0.1,10)$, $(1,100)$, $(2,200)$, and $(4,400)$.
$\omega_0$ is defined in the simulation frame.
We focus on the linear regime $a_0\omega_{pe}/\omega_0 \ll 1$, and fix $a_0\omega_{pe}/\omega_0=0.01$ throughout this study.
The initial wavenumber $k_0$ is numerically determined by the dispersion relation (Eq. \ref{eq:disp}).
The wavelength of the incident wave is resolved by 200 computational cells, $\lambda_0 = 200 \Delta x$. 
The time step is set as $\Delta t = \Delta x/c$. 
Our code employs an implicit Maxwell solver, which is not constrained by the CFL condition.
The size of simulation domain is set as $L_x=100\lambda_0$.
The initial spatial profiles of wave electric field $E_y = E_0y(\phi=-k_0x)$ and magnetic field 
$B_z  = \beta_gE_y$ are numerically determined by the self-consistent 
equation, Eq. \ref{eq:y}.

Particle motion must be consistent with the wave fields.
Eqs. \ref{eq:gam}, \ref{eq:ux}, and \ref{eq:uy}
indicate that the bulk Lorentz factor $\bar{\gamma}$ and bulk four velocity
$\bm{\bar{u}}$ at $t=0$ satisfy
\begin{eqnarray}
	\bar{\gamma} &=& 1+\frac{\alpha a_0^2}{2} \{1-[y(\phi=-k_0x)]^2\}, \\
	\bar{u}_{x} &=& \frac{\alpha \beta_g a_0^2}{2} \{1-[y(\phi=-k_0x)]^2\}, \\
	\bar{u}_{y} &=& a_0 \int_0^{-k_0x} y {\rm d}\phi, \\
	\bar{u}_{z} &=& 0.
\end{eqnarray}
We generate Maxwellian particles with a thermal spread $\beta_{th0}$ in the plasma rest frame,
and then compute the particle Lorentz factors and four velocities
in the simulation frame by performing the inverse Lorentz transformation.
We carry out our simulations for both strong- and weak-coupling cases:
$\beta_{th0}=0.01$ and $0.1$, respectively.
As described in Section \ref{subsec:sbs}, increasing the thermal velocity moves the system
from the strong-coupling regime to the weak-coupling regime. The weak-coupling
regime is particularly sensitive to kinetic effects, including Landau damping.
The growth rates and corresponding wavenumbers used in our comparison are
derived from a fully kinetic treatment of SBS and therefore include
finite-temperature effects. The fully kinetic PIC method also self-consistently
captures these kinetic effects in the numerical results.

The Debye length in units of $\Delta x$ is
$\lambda_D/\Delta x=(100/\pi)\beta_{th0}(\omega_0/\omega_{pe})$,
which ranges from $\simeq1.6$ to $\simeq1273$ for the parameters used here.
Thus, the stability condition $\Delta x \lesssim \lambda_D$ is satisfied. The
same numerical resolution was tested in Ref. \cite{Iwamoto2023}, where SBS in
pair plasmas was shown to be accurately captured.

\begin{figure}[htbp]
	\includegraphics[width=8cm]{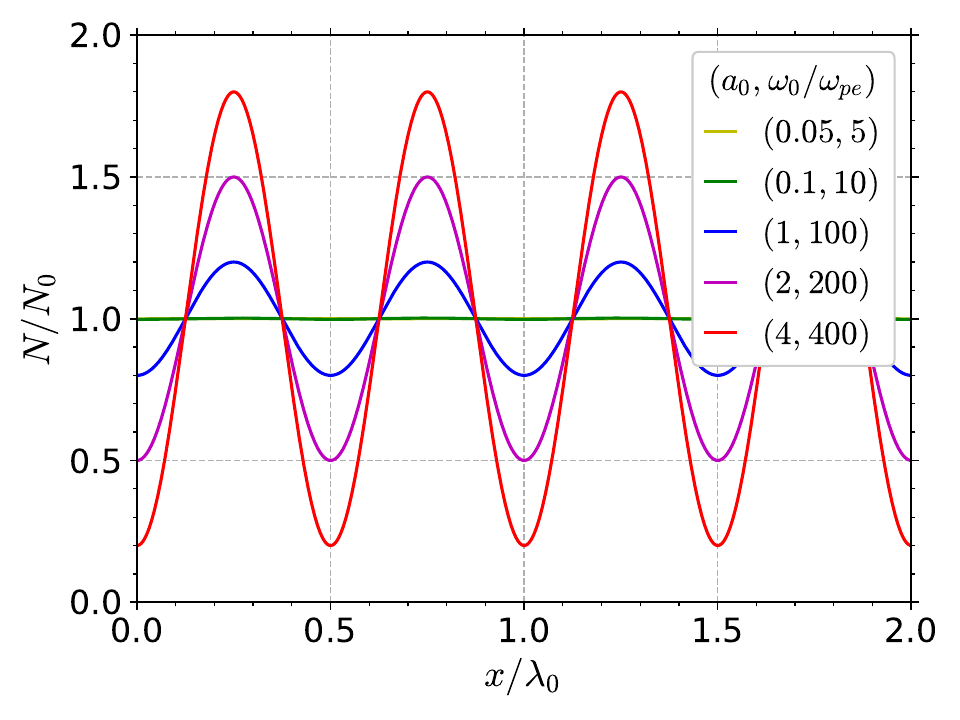}
	\caption{Enlarged view of the initial spatial profile of laboratory density $N$, normalized by the average
	density $N_0$, for various combinations of $(a_0,\omega_0/\omega_{pe})$. 
	The case $(a_0,\omega_0/\omega_{pe})=(0.05,5)$ (yellow) is overlapped by $(0.1,10)$ (green) and indistinguishable in the plot.}
	\label{fig:N}
\end{figure}

The particle position in the simulation frame is determined by the laboratory density $N$, 
which is related to the incident electric field as (see Eqs. \ref{eq:gam} and \ref{eq:n}),
\begin{equation}
	N = \gamma n = \frac{1 + \alpha a_0^2(1-y^2)/2}{1+2(1-y^2)/q} n_0.
\end{equation}
Since the laboratory density depends on both $a_0$ and $\omega_0/\omega_{pe}$, 
we evaluate it for each set of parameters.
Fig. \ref{fig:N} shows an enlarged view of the initial spatial profile of the laboratory density,
normalized by the averaged density $N_0=\langle N \rangle$.
For $a_0\omega_{pe}/\omega_0 \ll 1$ (i.e., $q \gg 1$), one can find
\begin{equation}
	N_0 \simeq \left(1+\frac{a_0^2}{4}\right)n_0
\end{equation}
for the zeroth-order of $a_0\omega_{pe}/\omega_0$.
We distribute particle positions accordingly to represent the density profile.
To ensure adequate resolution of the density fluctuations,
the averaged number of particles per species per cell is set to $N_0\Delta x = 100$.

\subsection{\label{subsec:linear} Linear Stage}
\begin{figure}[htbp]
	\includegraphics[width=8.5cm]{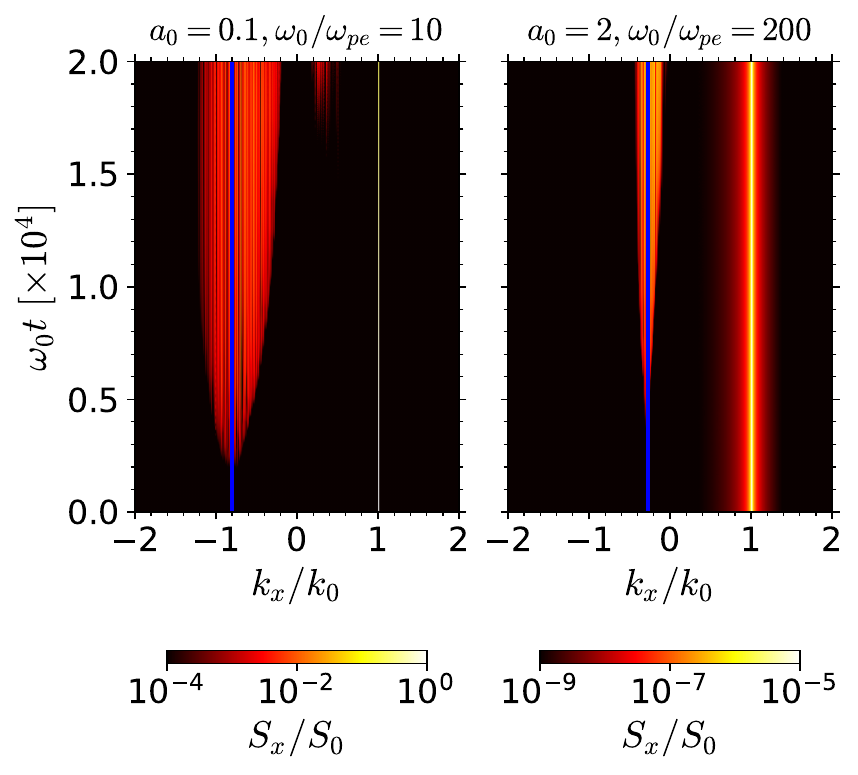}
	\caption{Time evolution of the power spectrum of the Poynting flux for $(a_0,\omega_0/\omega_{pe})=(0.1,10)$ (left)
	and $(2,200)$ (right). The initial thermal velocity in the plasma rest frame is $\beta_{th0}=0.1$ in both cases.
	The blue lines represent the theoretical fastest-growing modes (Eq. \ref{eq:klab1}).}
	\label{fig:spectra}
\end{figure}

We first focus on the linear stage of the wave evolution and verify our assumption in Section \ref{subsec:sbs}.
Fig. \ref{fig:spectra} shows the time evolution of the power spectrum of the Poynting flux $S_x$
normalized by its initial value $S_{0}$ for $(a_0,\omega_0/\omega_{pe})=(0.1,10)$ (left) and $(2,200)$ (right).
The initial thermal velocity in the proper frame is $\beta_{th0}=0.1$ (weak coupling) in both cases.
We perform a spatial Fourier transform of $S_x$ at each snapshot and 
distinguish between oppositely propagating wave components ($k_x > 0$ and $k_x < 0$), 
following the method described in Ref. \cite{Ghosh2022}.
The back-scattered waves ($k_x < 0$) are generated by SBS for both cases,
and the absolute wavenumber of the fastest-growing mode for $a_0 = 2$ (left) is smaller than that for 
$a_0 = 0.1$ (right), which is due to the Lorentz boost effect as discussed in Section \ref{subsec:sbs}.
The theoretical prediction given by Eq. \ref{eq:klab1} (blue lines) is 
indeed consistent with the simulation results.

\begin{figure*}[htbp]
	\includegraphics[width=16cm]{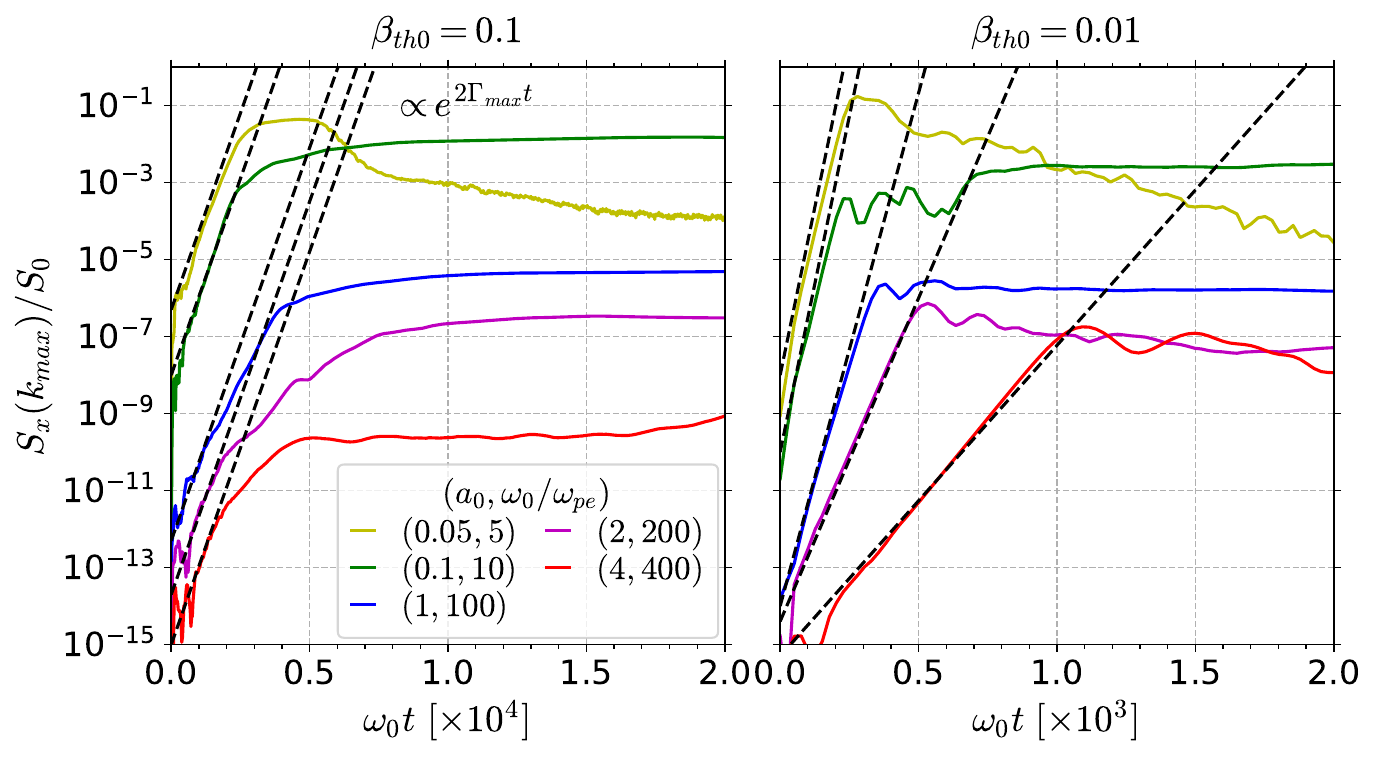}
	\caption{Time evolution of the Poynting flux $S_x(k_{\mathrm{max}})$ associated with the fastest-growing mode 
	for $\beta_{th0}=0.1$ (left) and $\beta_{th0}=0.01$ (right) for various combinations of $(a_0,\omega_0/\omega_{pe})$. 
	The black dashed lines represent $\propto e^{2\Gamma_{\mathrm{max}}t}$, where $\Gamma_{\mathrm{max}}$ is the theoretical maximum growth rate 
	in the simulation frame (Eqs. \ref{eq:gamlab1} and \ref{eq:gamlab2}).}
	\label{fig:sx}
\end{figure*}

Fig. \ref{fig:sx} shows the time evolution of the Poynting 
flux $S_x$ in the simulation frame associated with the fastest-growing mode $k_\mathrm{max}$ for 
$\beta_{th0} = 0.1$ (left) and $\beta_{th0} = 0.01$ (right), for various combinations of 
$(a_0,\omega_0/\omega_{pe})$: $(0.05,5)$ (yellow), $(0.1,10)$ (green), $(1,100)$ (blue), 
$(2,200)$ (magenta), and $(4,400)$ (red).
The maximum growth rates for $\beta_{th0}=0.1$ are independent of $a_0$, whereas
those for $\beta_{th0}=0.01$ decrease with increasing $a_0$.
The black dashed lines represent exponential growth with the theoretical maximum growth rate $\propto e^{2\Gamma_{\mathrm{max}}t}$, where $\Gamma_{\mathrm{max}}$ is 
calculated from Eqs. \ref{eq:gamlab1} and \ref{eq:gamlab2}.
The linear growth rates are well reproduced by the theoretical predictions for both $\beta_{th0}$.
We also checked the density perturbation as a consistency check. The density
spectrum shows power around the wavenumber expected from the matching
condition, $k_{\rm SBS}=k_0-k_{1,\max}$.
On the other hand, the saturation levels vary significantly depending on the combination of $(a_0,\omega_0/\omega_{pe})$.
The saturation behavior will be discussed in detail in the next section.

\begin{figure}[htbp]
	\includegraphics[width=8cm]{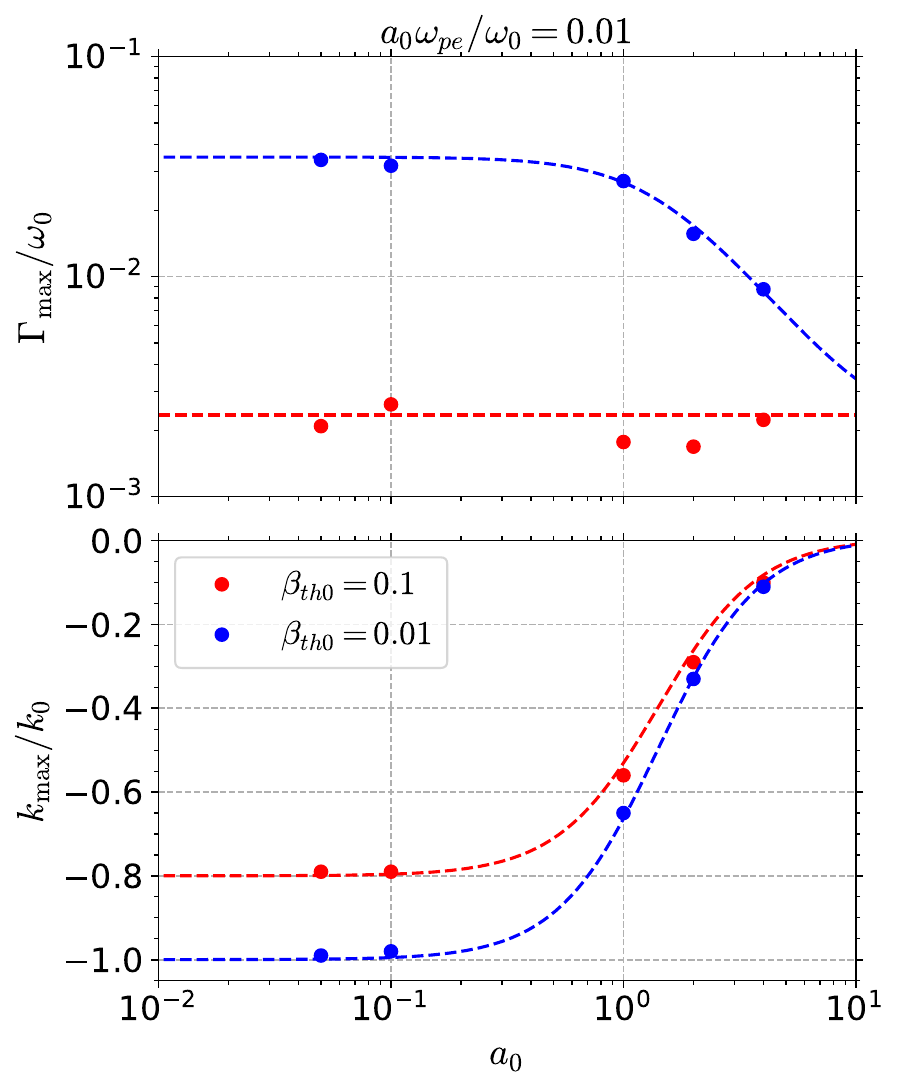}
	\caption{Maximum growth rate (top) and corresponding wavenumber (bottom) of the scattered wave as a function of $a_0$
	while keeping the nonlinearity parameter $a_0\omega_{pe}/\omega_0=0.01$. The color represents the results for $\beta_{th0}=0.1$ (red) and $\beta_{th0}=0.01$ (blue).
	The theoretical predictions from Eqs. \ref{eq:gamlab1},  \ref{eq:klab1}, \ref{eq:gamlab2},and \ref{eq:klab2} are shown by the dashed lines.}
	\label{fig:growth}
\end{figure}

Figure \ref{fig:growth} shows the maximum growth rate (top) and the corresponding wavenumber (bottom) of the scattered wave as a function of $a_0$ for 
$\beta_{th0}=0.1$ (red circles) and $\beta_{th0}=0.01$ (blue circles). 
In the weak coupling regime ($\beta_{th0}=0.1$), the maximum growth rates are independent of $a_0$, whereas they decrease with increasing $a_0$ in the strong coupling regime 
($\beta_{th0}=0.01$). In both cases, the wavenumbers of the backscattered wave approach zero as $a_0$ increases.
These behaviors are well explained by the theoretical predictions 
from Eqs. \ref{eq:gamlab1} and \ref{eq:klab1} for the weak coupling regime ($\beta_{th0} = 0.1$), and 
Eqs.  \ref{eq:gamlab2} and \ref{eq:klab2} for the strong coupling regime ($\beta_{th0} = 0.01$).
The theoretical predictions are represented by dashed lines in the corresponding colors, and they are in good agreement with the simulation results.
The agreement between the simulation results and theoretical predictions confirms that the $a_0$ dependence of the growth rate and wavenumber is 
due to the Lorentz boost effect. This result indicates that the linear analysis of SBS is extrapolated to the regime $a_0 > 1$ as long as the nonlinearity parameter is sufficiently 
small, $a_0\omega_{pe}/\omega_0 \ll 1$.

A recent related study of strong electromagnetic waves in unmagnetized pair plasmas 
found a scattered-wavenumber scaling consistent with our result, while the growth-rate 
behavior is treated in a different setup \cite{Sridhar2026}. In particular, the 
related study considers wave packets rather than plane waves, and does not 
address the weak-coupling regime considered here. Developing a comprehensive theory 
that connects these complementary cases is an important subject for future work.

\subsection{\label{subsec:nonlinear} Nonlinear Stage}
\begin{figure*}[htbp]
	\includegraphics[width=16cm]{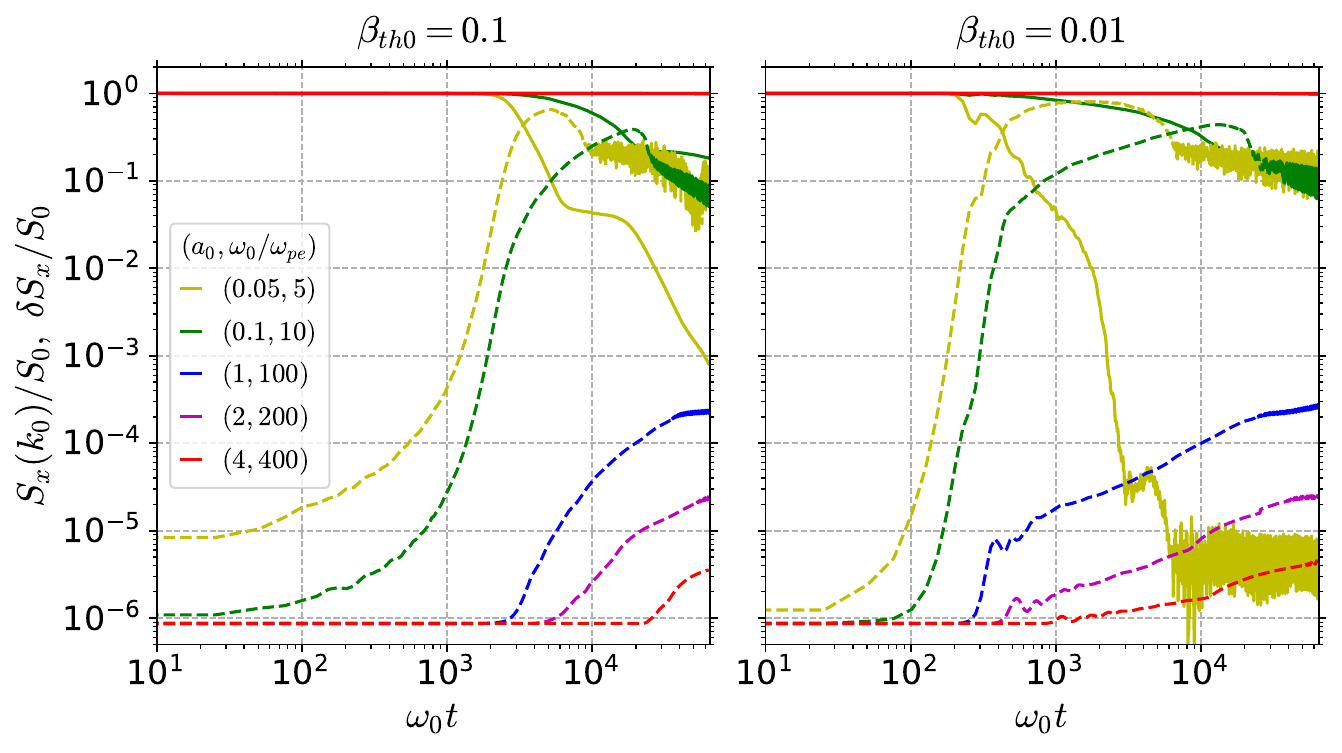}
	\caption{Time evolution of the incident Poynting flux $S_x(k_0)$ (solid lines) and scattered Poynting flux $\delta S_x$ (dashed lines)
	for $\beta_{th0}=0.1$ (left) and $\beta_{th0}=0.01$ (right). The color represents the results for various combinations of $(a_0,\omega_0/\omega_{pe})$ 
	in the same manner as Fig. \ref{fig:sx}.}
	\label{fig:dsx}
\end{figure*}

We now discuss the nonlinear stage of SBS. The saturation levels of the fastest-growing modes (Fig. \ref{fig:sx}) tend to decrease with increasing $a_0$ while keeping
the nonlinearity parameter $a_0\omega_{pe}/\omega_0$ constant for both $\beta_{th0}$, indicating that the incident waves are less affected by SBS for larger $a_0$. 
Fig. \ref{fig:dsx} shows the time evolution of the incident Poynting flux $S_x(k_0)$ (solid lines) and scattered Poynting flux $\delta S_x$ (dashed lines) 
for $\beta_{th0}=0.1$ (left) and $\beta_{th0}=0.01$ (right). The scattered Poynting flux $\delta S_x$ is defined as the sum of the Poynting fluxes over all scattered modes, 
\begin{equation}
	\delta S_x = \sum_{k_x \neq k_0} S_x(k_x).
\end{equation}
The color represents the results for various combinations of $(a_0,\omega_0/\omega_{pe})$ , following
the same convention as in Fig. \ref{fig:sx}. For $(a_0,\omega_0/\omega_{pe})$ = $(0.05,5)$ (yellow) and $(0.1,10)$ (green), 
a significant fraction of the incident Poynting flux is dissipated, and the scattered Poynting flux becomes comparable to the incident 
flux by the end of the simulation for both values of $\beta_{th0}$. As $a_0$ increases, the incident Poynting flux is less affected by SBS, 
and the scattered Poynting flux decreases. This behavior can be explained as follows.

The incident wave energy is dissipated via Landau damping of the acoustic-like modes and subsequently converted into plasma kinetic energy. When the incident wave energy is 
sufficiently large, it requires a considerable amount of time to transfer a significant fraction of this energy to the plasma. The ratio of the incident wave energy density to the 
electron rest mass energy density is given by
\begin{equation}
	\frac{E_0^2}{4\pi n_0 m_e c^2} = \left( a_0 \frac{\omega_0}{\omega_{pe}} \right)^2.
\end{equation}
For the cases $(a_0, \omega_0/\omega_{pe}) = (0.05, 5)$ (yellow) and $(0.1, 10)$ (green), this ratio is smaller than or comparable to unity. Consequently, the incident wave energy 
is dissipated relatively quickly, 
and SBS effects become significant within the simulation timescale. In contrast, for $(a_0, \omega_0/\omega_{pe}) = (1, 100)$ (blue), $(2, 200)$ (red), and 
$(4, 400)$ (magenta), this ratio is much larger than unity, meaning that the incident wave energy is hardly converted into plasma energy during the simulation. This argument 
qualitatively explains the dependence of the saturation levels; thus, the energy ratio $a_0 \omega_0 / \omega_{pe}$ likely governs the saturation behavior of SBS.
Note a different combination of parameters from the nonlinearity parameter $a_0 \omega_{pe}/\omega_0$.

\begin{figure*}[htbp]
	\includegraphics[width=17cm]{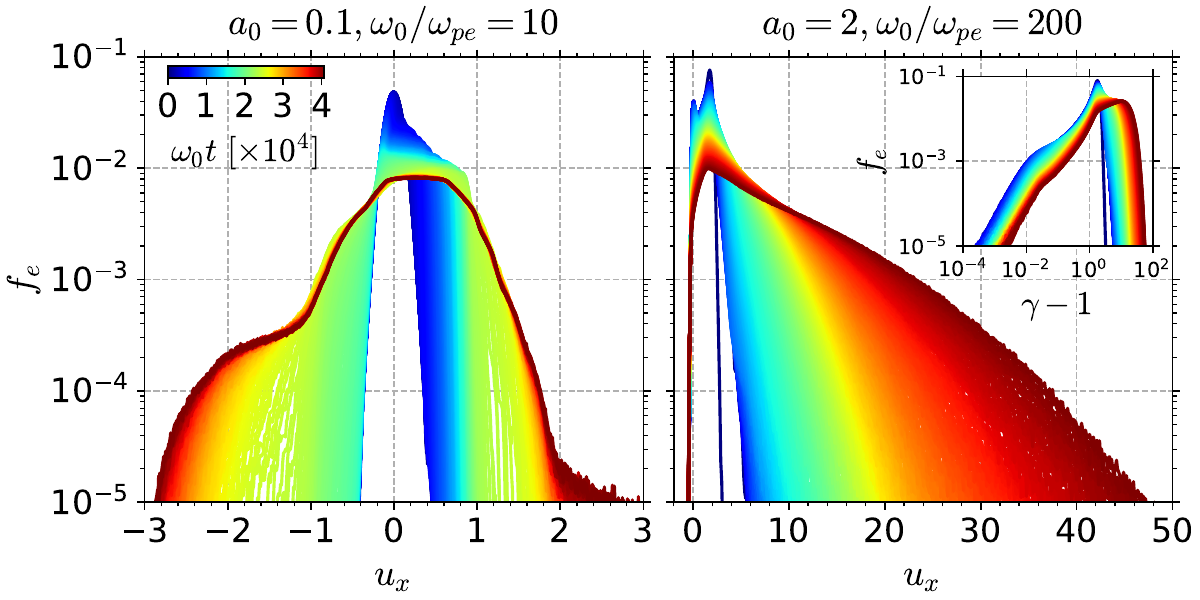}
	\caption{Time evolution of the longitudinal four velocity $u_x$ distribution in the simulation frame for $(a_0,\omega_0/\omega_{pe}) = (0.1,10)$ (left) and $(2,200)$ (right) with $\beta_{th0}=0.1$ (weak coupling). The inset in the right panel displays the corresponding particle energy spectra on a log-log scale.}
	\label{fig:dist}
\end{figure*}

Although the incident wave is less affected by SBS at larger ratio $a_0 \omega_0/\omega_{pe}$, the plasma is more significantly modified in this regime. 
This occurs because, when wave 
energy dominates ($a_0\omega_0/\omega_{pe} \gg 1$), even the dissipation of a small fraction of the incident energy is sufficient to perturb the velocity distribution 
substantially. Figure \ref{fig:dist} shows the time evolution of the longitudinal four-velocity $u_x$ distribution in the simulation frame for 
$(a_0, \omega_0/\omega_{pe}) = (0.1, 10)$ (left) and $(2, 200)$ (right) with $\beta_{th0}=0.1$. Landau damping of the acoustic-like modes excited by SBS results in 
the formation of a plateau in the $u_x$ distribution, which is a characteristic feature of SBS-induced heating reported in previous studies \cite{Matsukiyo2003}.
The development of such a plateau triggers the saturation of SBS because the resonant coupling, which depends on the gradient of the velocity distribution function, 
is reduced as the distribution flattens \cite{Kamijima2026, Nishiura2026}.
For the $(a_0, \omega_0/\omega_{pe}) = (0.1, 10)$ case, a distinct plateau forms at early stages around $u_x \sim 0.1$, which is comparable to the initial thermal velocity 
$\beta_{th0}=0.1$, as in the previous studies \cite{Kamijima2026,Nishiura2026}.
This plateau subsequently expands toward both larger and smaller $u_x$. In the nonlinear stage, scattered waves trigger secondary SBS, further modifying the 
distribution. 

In contrast, for the $(a_0, \omega_0/\omega_{pe}) =(2, 200)$ case, the initial velocity distribution is heavily influenced by the incident wave; consequently, a standard 
Maxwellian distribution 
is not observed in the simulation frame even at $\omega_0 t=0$. While particles are initialized with a Maxwellian distribution ($\beta_{th0}=0.1$) in the plasma rest frame, the 
large-amplitude incident wave ($a_0 > 1$) significantly shifts the distribution in the simulation frame (see Eq. \ref{eq:ux}). Although the Landau damping process is more complex 
in this case because of the relativistic oscillatory motion of the particles, the $u_x$ distribution is nonetheless significantly modified by SBS, with the high-energy 
tail extending to extremely large $u_x$ at later times. The inset in the right panel of Fig. \ref{fig:dist} shows the corresponding particle energy spectra on a log-log scale, 
demonstrating significant energy gain. However, a clear power-law distribution or a distinct hot component is not observed by the end of the simulation under the present physical 
conditions (cf. Refs. \cite{Matsukiyo2009, Sano2024, Isayama2025}). These results demonstrate that particle heating becomes more pronounced as $a_0\omega_0/\omega_{pe}$ increases, 
even when the backreaction on the incident wave remains small.

\section{\label{sec:discuss} Discussion}
We discuss the implications of our results for FRBs. The radio pulses propagate through the magnetar wind,
and the laboratory (simulation) frame in our analysis corresponds to the frame where the magnetar wind is at rest prior to the arrival of the pulse.
Note that the large-amplitude incident waves drive the plasma in the wave propagation direction (see Eq. \ref{eq:vD}).
Given that the typical frequency of FRBs is $\nu_{\mathrm{obs}} \sim 1$ GHz in the observer frame \cite{Petroff2022},
the incident wave frequency in the wind rest frame is given by
\begin{equation}
	\label{eq:w_frb}
	\omega_0 = \frac{\pi \nu_{\mathrm{obs}}}{\gamma_{\mathrm{wind}}} \sim  3 \times 10^7 \ \nu_{\mathrm{obs}, 9} \gamma_{\mathrm{wind}, 2}^{-1} \ \mathrm{rad \ s^{-1}},
\end{equation}
where $\gamma_{\mathrm{wind}} \sim 10^2$ is the Lorentz factor of the magnetar wind \cite{Beloborodov2020}, $\nu_{\mathrm{obs}, 9} = \nu_{\mathrm{obs}}/10^9$ Hz, 
and $\gamma_{\mathrm{wind}, 2} = \gamma_{\mathrm{wind}}/10^2$. Hereafter, we use the notation $Q_x = Q/10^x$.
The isotropic radio luminosity of FRBs is $L_{\mathrm{obs}} = 2c\gamma_{\mathrm{wind}}^2 E_0^2 R^2 \sim 10^{42}$ erg s$^{-1}$,
where $R$ is the distance from the magnetar, and thus the strength parameter is estimated as
\begin{equation}
	\label{eq:a_frb}
	a_0 \sim 20 L^{\frac{1}{2}}_{\mathrm{obs},42} \nu^{-1}_{\mathrm{obs}, 9} R^{-1}_{12}.
\end{equation}
The electron number density in the wind rest frame is 
\begin{equation}
	n_0 = \frac{\dot{N}}{8\pi \gamma_{\mathrm{wind}} R^2 c} \sim 10 \ \dot{N}_{39} \gamma_{\mathrm{wind}, 2}^{-1} R^{-2}_{12} \ \mathrm{cm^{-3}},
\end{equation}
where $\dot{N} \sim 10^{39}$ s$^{-1}$ is the particle flux of the magnetar wind \cite{Beloborodov2020}.
Note that $\dot{N}$ is still uncertain. The electron plasma frequency in the wind rest frame is then given by
\begin{equation}
	\omega_{pe} \sim 2 \times 10^5 \ \dot{N}^{\frac{1}{2}}_{39} \gamma_{\mathrm{wind}, 2}^{-\frac{1}{2}} R^{-1}_{12} \ \mathrm{rad \ s^{-1}}.
\end{equation}
The nonlinearity parameter is estimated as
\begin{equation}
	\label{eq:nonlinearity}
	a_0 \frac{\omega_{pe}}{\omega_0} \sim 10^{-1} \ L^{\frac{1}{2}}_{\mathrm{obs},42} \dot{N}^{\frac{1}{2}}_{39} \nu^{-2}_{\mathrm{obs}, 9} \gamma_{\mathrm{wind}, 2}^{\frac{1}{2}} R^{-2}_{12}.
\end{equation}
This indicates that linear SBS analysis remains applicable for $R \sim 10^{12}$ cm despite the large amplitude $a_0 > 1$.
Assuming that the thermal velocity of the magnetar wind is controlled by adiabatic expansion and Compton heating by X-rays emitted from the magnetar \cite{Sobacchi2023},
we estimate
\begin{equation}
	\beta_{th0} \sim 10^{-3} R_{12}^{-\frac{1}{2}}.
\end{equation}
Thus, the strong coupling regime is relevant at $R \sim 10^{12}$ cm, where the plasma temperature does not significantly affect the SBS growth rate or wavenumber.
The linear growth timescale of SBS is given by
\begin{equation}
	\tau_{\mathrm{SBS}} = \frac{1}{\Gamma_{\mathrm{max}}} \sim 3 \times 10^{-6} L^{\frac{1}{3}}_{\mathrm{obs},42}\dot{N}^{-\frac{1}{3}}_{39} \nu^{-1}_{\mathrm{obs}, 9} \gamma_{\mathrm{wind}, 2}^{\frac{2}{3}}\ \mathrm{s},
\end{equation}
We use Eqs. \ref{eq:gamlab2}, \ref{eq:gam_D}, \ref{eq:w_frb}, \ref{eq:a_frb}, and \ref{eq:nonlinearity} to derive the above estimate.
We compare $\tau_{\mathrm{SBS}}$ with the time duration of the characteristic timescale of FRBs.
The time duration of the radio pulse in the wind rest frame $\tau_{\mathrm{pulse}}$ is
\begin{equation}
  \tau_{\mathrm{pulse}} = 2\gamma_{\mathrm{wind}}\tau_{\mathrm{obs}} \sim 2 \times 10^{-1} \gamma_{\mathrm{wind}, 2} \tau_{\mathrm{obs}, -3} \ \mathrm{s},
\end{equation}
where $\tau_{\mathrm{obs}} \sim 10^{-3}$ s is the observed pulse duration.
The dynamical time in the wind rest frame is
\begin{equation}
	\tau_{\mathrm{dyn}} = \frac{R}{2\gamma_{\mathrm{wind}}c} \sim 2 \times 10^{-1} \ R_{12} \gamma_{\mathrm{wind}, 2}^{-1} \ \mathrm{s}.
\end{equation}
Since $\tau_{\mathrm{pulse}} \lesssim \tau_{\mathrm{dyn}}$ for $R \gtrsim 10^{12}$ cm, the relevant timescale for the growth of SBS during the FRB propagation 
is $\tau_{\mathrm{pulse}}$. The pulse duration is much longer than the linear growth timescale: $\tau_{\mathrm{pulse}} \gg \tau_{\mathrm{SBS}}$, 
indicating that SBS can grow during the FRB propagation in the magnetar wind.

Regarding nonlinear evolution, the saturation level of SBS is characterized by the ratio of the incident wave energy to 
the rest mass energy, $a_0\omega_0/\omega_{pe}$ as discussed in Section \ref{subsec:nonlinear}.
This ratio for FRBs is estimated as
\begin{equation}
	a_0 \frac{\omega_0}{\omega_{pe}} \sim 3 \times 10^3 \ L^{\frac{1}{2}}_{\mathrm{obs},42} \dot{N}^{-\frac{1}{2}}_{39} \gamma_{\mathrm{wind}, 2}^{-\frac{1}{2}},
\end{equation}
which is comparable to the largest value in our simulations, $a_0\omega_0/\omega_{pe} = 1600$.
For such a large value of $a_0\omega_0/\omega_{pe}$, the incident wave is barely affected by SBS 
by the end of our simulation,
\begin{equation}
	\omega_0 \tau_{\mathrm{end}} \sim 6 \times 10^4.
\end{equation}
On the other hand, the characteristic FRB timescale in units of $\omega_0$ is given by
\begin{equation}
	\omega_0\tau_{\mathrm{pulse}} \sim 6 \times 10^6 \ \nu_{\mathrm{obs}, 9} \tau_{\mathrm{obs}, -3}.
\end{equation}
Although $\omega_0\tau_{\mathrm{pulse}}$ significantly exceeds the simulation duration $\omega_0 \tau_{\mathrm{end}}$, several factors likely suppress the dissipation rate in 
realistic astrophysical scenarios. First, our periodic boundary conditions prevent scattered waves from escaping the interaction region, which may lead to an artificial enhancement 
of SBS growth and energy dissipation. In actual FRB propagation, scattered waves naturally escape from the pulse region, while the leading edge of the incident wave continuously 
encounters unperturbed plasma. Consequently, the growth of SBS is limited by the convection of scattered waves, thereby reducing the net dissipation rate of the incident pulse. 
We plan to address these propagation effects in future work using open-boundary simulations, following Ref. \cite{Sridhar2026}.
Second, SBS is known to be suppressed for broadband incident waves \cite{Ghosh2022, Nishiura2025a}, a state characteristic of FRB signals. 
Furthermore, the filamentation instability \cite{Sobacchi2020, Sobacchi2022, Sobacchi2023}, which is not captured in our current 1D model, typically grows faster than 
SBS \cite{Ghosh2022}. Since filamentation is relatively insensitive to Landau damping \cite{Iwamoto2023}, it may further inhibit the SBS energy loss channel. 
In addition, as plasma heating progresses, the system may transition from a strong coupling to a weak coupling regime, where the SBS growth rate is significantly lower. 
Finally, for $a_0 \gg 1$, the ponderomotive force can trigger bulk acceleration of the magnetar wind. An increase in the wind Lorentz factor $\gamma_{\mathrm{wind}}$ effectively 
shortens the dynamical timescale $\tau_{\mathrm{dyn}}$ in the wind rest frame, such that the SBS cannot grow sufficiently during the FRB propagation.
Therefore, we conclude that FRBs can propagate through the magnetar wind at $R \gtrsim 10^{12}$ cm without substantial energy 
loss for our fiducial parameters, even if the signals undergo spectral or temporal modifications.

\section{\label{sec:summary} Summary}
We have investigated the induced scattering of linearly-polarized, large-amplitude electromagnetic waves in unmagnetized pair plasmas using analytical theory and PIC simulations.
Our results demonstrate that  the steady-state solution is governed by the nonlinearity parameter $a_0 \omega_{pe} / \omega_0$ rather than $a_0$ itself. 
In the regime where $a_0 \omega_{pe} / \omega_0 \ll 1$, the plasma current follows the test-particle limit, allowing the solution
to remain essentially linear even for $a_0 > 1$.

We showed that the linear growth rate and wavenumber of SBS in the simulation frame depend on both $a_0 \omega_{pe} / \omega_0$ and $a_0$, a consequence of the Lorentz boost from 
the incident-wave-driven bulk motion. PIC simulations confirm that conventional linear theory can be extrapolated to the $a_0 > 1$ regime, provided the nonlinearity parameter 
remains small. Furthermore, the SBS saturation level is found to be controlled by the energy ratio $a_0 \omega_0 / \omega_{pe}$. When $a_0 \omega_0 / \omega_{pe} \gg 1$, incident 
wave dissipation is minimal despite significant particle heating via Landau damping.

Applying these results to FRBs in magnetar winds, we find that $a_0 \sim 20$ and $a_0 \omega_{pe} / \omega_0 \sim 10^{-1}$ at $R \sim 10^{12}$ cm
for our fiducial parameters.
These values indicate that linear SBS analysis remains applicable for FRBs, and the growth timescale of SBS is much shorter than the pulse duration. However, the saturation level 
of SBS is expected to be low due to the large value of $a_0 \omega_0 / \omega_{pe} \sim 10^3$, suggesting that FRBs can propagate through the magnetar wind without substantial 
energy loss, even if they undergo spectral or temporal modifications.

In this work, we neglect the effects of the background magnetic field, an approximation that remains valid in regions far from the magnetar. In highly magnetized plasmas
$\omega_L \gg \omega_{pe}$, where $\omega_L$ is the cyclotron frequency, the nonlinearity parameter is effectively defined by $a_0 \omega_L / \omega_0$ rather than 
$a_0 \omega_{pe} / \omega_0$ \cite{Sobacchi2024b,Sobacchi2025}. Under the condition $a_0 \omega_L / \omega_0 \ll 1$, the incident wave can be treated within a linear framework; 
consequently, linear analyses of induced scattering in magnetized pair plasmas \cite{Nishiura2025a,Nishiura2025b,Kamijima2026,Nishiura2026} 
might remain applicable even for $a_0 > 1$. The influence of the background 
magnetic field on the propagation of large-amplitude waves will be investigated in a future publication.

\acknowledgments

We are grateful to  W. Ishizaki,  S. F. Kamijima, P. Kumar, R. Kuze, R. Nishiura, K. Sugimoto, and Y. Takei for fruitful discussions.
MI thanks E. Sobacchi, L. Sironi, N. Sridhar, and D. Groselj for helpful conversations.
We thank the Yukawa Institute for Theoretical Physics at Kyoto University. Discussions during the YITP long-term workshop YITP-T-26-02 on ``Multi-Messenger Astrophysics in the Dynamic Universe'' were useful to complete this work.
We acknowledge support from JSPS KAKENHI Grant No. 22H00130.
MI acknowledges support from JSPS KAKENHI Grants No. 23K20038, No. 26K17150, and No. 26K00697.
KI acknowledges support from JSPS KAKENHI Grant No. 23H04900, 23H05430, 23H01172, and
26H02045.
This work was supported by MEXT as ``Program for Promoting Researches on the Supercomputer Fugaku''
(Structure and Evolution of the Universe Unraveled by Fusion of Simulation and AI; 
Grant Number JPMXP1020230406) and used computational resources of supercomputer Fugaku provided 
by the RIKEN Center for Computational Science (Project ID: hp240182, hp240219, hp250161, hp250226,).
This work used the computational resources of the HPCI system provided by 
Information Technology Center, Nagoya University, through the HPCI System 
Research Project (Project ID: hp240147, hp250036).

\bibliography{ref}
\end{document}